\documentclass[3p,times,twocolumn,authoryear]{elsarticle}

\usepackage{ecrc}
\usepackage[usenames, dvipsnames]{color}
\usepackage{breqn}
\usepackage{bm}
\usepackage{soul}
\definecolor{darkgreen}{rgb}{0.01, 0.75, 0.24}

\volume{00}

\firstpage{0}

\journalname{submitted to Planetary and Space Science}

\runauth{E. Chan\'e et al.}

\usepackage{amssymb}

\usepackage[figuresright]{rotating}

\begin{document}

\begin{frontmatter}



\dochead{}

\title{Periodic shearing motions in the Jovian magnetosphere causing a localized peak in the main auroral emission close to noon.}


\author[1]{E. Chan\'e}

\author[2]{B. Palmaerts}

\author[2]{A. Radioti}

\address[1]{Centre for mathematical Plasma Astrophysics, KU Leuven, Leuven, Belgium.}

\address[2]{Laboratoire de Physique Atmosph\'erique et Plan\'etaire, STAR Institute, Universit\'e de Li\`ege, Li\`ege, Belgium.}

\begin{abstract}
Recently, a transient localized brightness enhancement has been observed in Jupiter's main auroral emission close to noon by \citet{Palmaerts2014}.
We use results from three-dimensional global MHD simulations to understand what is causing this localized peak in the main emission.
In the simulations, the peak occurs every rotation period and is due to shearing motions in the magnetodisk. 
These shearing motions are caused by heavy flux-tubes being accelerated to large azimuthal speeds at dawn. The centrifugal force acting on these
flux-tubes is then so high that they rapidly move away from the planet. When they reach noon, their azimuthal velocity decreases, thus reducing the centrifugal force, 
and allowing the flux-tubes to move back closer to Jupiter.
The shearing motions associated with this periodic phenomenon locally increase the field aligned currents in the simulations, thus causing a transient brightness enhancement in the 
main auroral emission, similar to the one observed by \citet{Palmaerts2014}.
\end{abstract}


\end{frontmatter}


\begin{figure}
\begin{center}
\noindent\includegraphics[width=19pc]{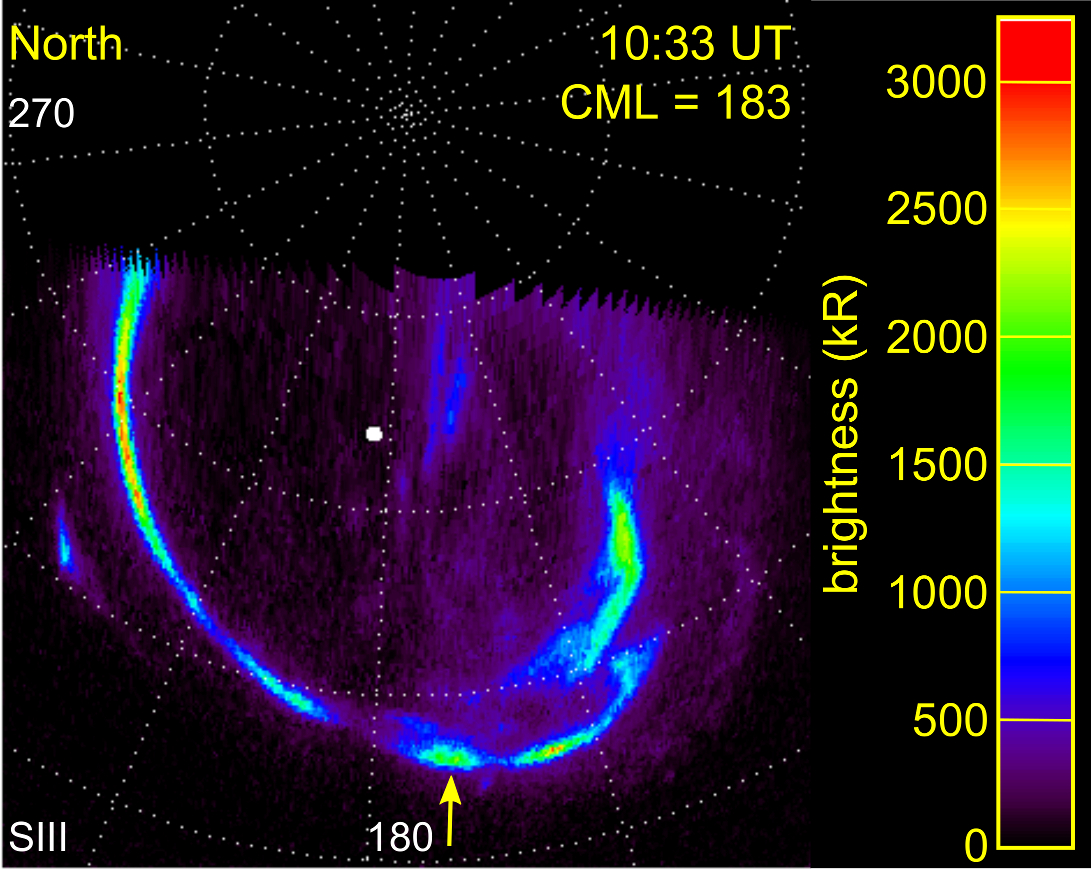}
\caption{
Polar projection in System~III polar coordinate system of an auroral image obtained with the Advanced Camera for Surveys (ACS) of the Hubble Space Telescope on February 7, 2006. 
System III 180$^\circ$ meridian and the magnetic local noon are oriented toward the bottom of the page. 
The transient localized auroral peak is indicated by the yellow arrow. 
Further information about this image and the whole sequence of observation can be found in \cite{Palmaerts2014} where this image comes from.
}
\label{fig:HST}
\end{center}
\end{figure}

\begin{figure*}
\begin{center}
\noindent\includegraphics[width=40pc]{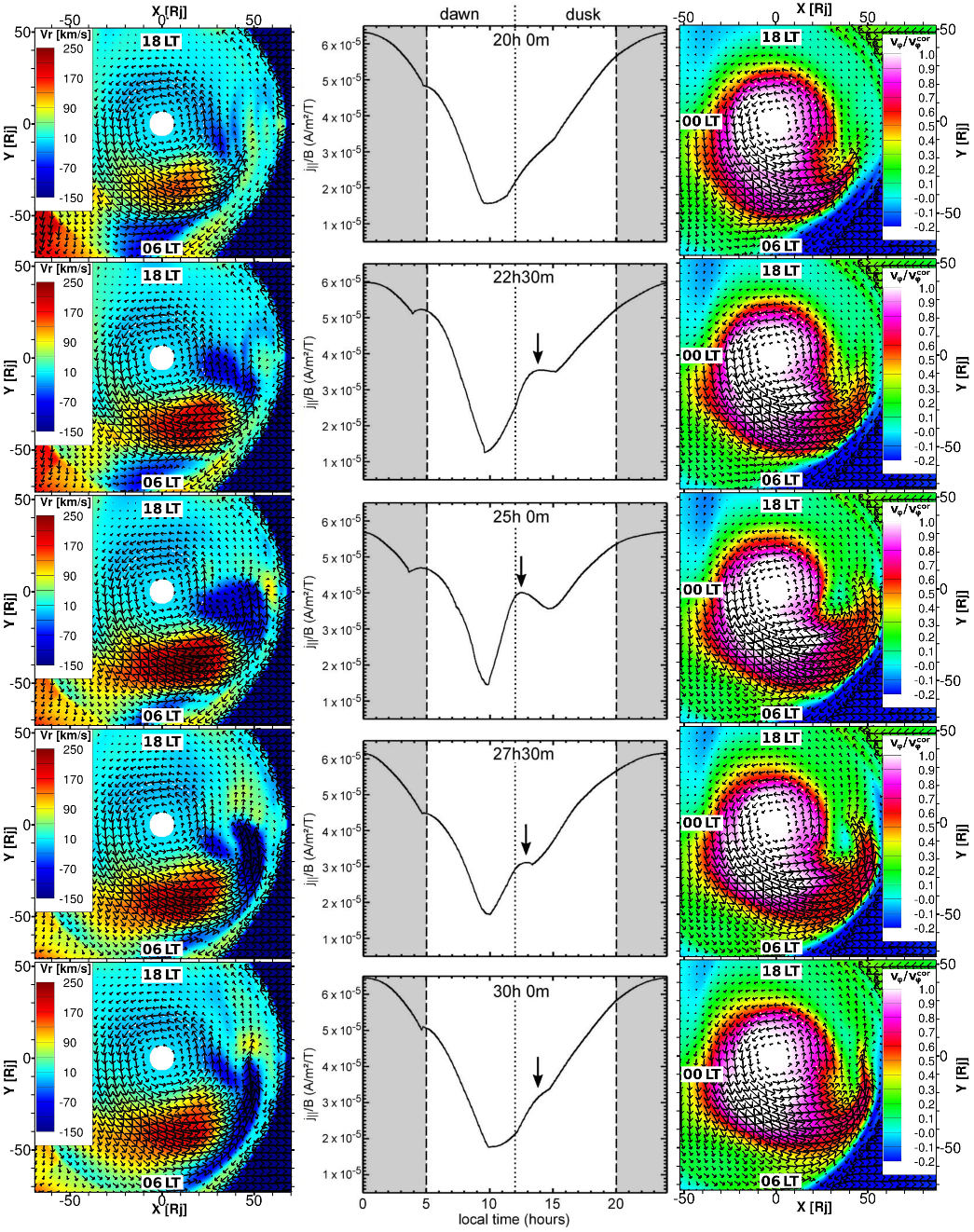}
\caption{Left panels: radial velocity (color coded) and velocity vectors in the equatorial plane.
Central panels: $j_{\parallel}/B$ in the ionosphere (peak value) as a function of local-time.
The localized peak near noon local-time on the right panels is associated with negative radial velocities 
(and thus shearing motions) on the left panels.
Right panels: the ratio between the plasma azimuthal velocity ($v_\phi$) and the theoretical corotation velocity
($v_\phi^{cor}$) and velocity vectors in the equatorial plane.
The color bars are saturated.
}
\label{fig:VrJpar}
\end{center}
\end{figure*}

\section{Introduction}
\label{sec:intro}
At Jupiter, the main auroral emission is located at about 15$^\circ$ magnetic colatitude on both hemispheres.
This very bright (it can reach intensities of MR) emission is continuously present,
relatively steady, and corotates with the planet (it is fixed in system III longitude).
The main emission is clearly visible in ultraviolet, in visible light, and in thermal infrared \citep[see][and references therein]{Clarke2004,Badman2015,Grodent2015}. 
\citet{Radioti2008}, using observations from the Hubble space telescope (HST), showed that the main emission displays a discontinuity fixed in magnetic local-time,
between 08:00~LT and 13:00~LT, where the intensity of the emission is lower.

The main auroral emission has been shown to be generated by the breakdown of corotation of the plasma located at a radial distance of about 25~R$_{\textrm{\scriptsize{J}}}$
in the magnetodisk. The large majority of the plasma inside Jupiter's magnetosphere originates from the volcanic moon Io, or from the neutral cloud associated with this moon.
Close to Jupiter, the Iogenic plasma rigidly corotates with the planet and slowly moves away to larger radial distances. Eventually, this plasma is ejected in the magnetotail 
of Jupiter.
While moving to larger radial distances, in the absence of external forces, the plasma angular velocity 
would decrease due to the conservation of angular momentum. 
This is not the case close to Jupiter, where strong magnetic Lorentz forces accelerate the plasma up to rigid corotation.
At a given radial distance ---that might be time-dependent and that can also depend on local-time--- the Lorentz forces are not strong enough to accelerate the plasma up to 
rigid corotation: the plasma starts to sub-corotate. This is called the breakdown of corotation and it affects the shape of the magnetic field lines, which begin to bent
in the azimuthal direction (since the feet of the field lines rotate faster than their appexes).
This bent back of the field lines produces a radial electrical current, which is closed via an equatorward Pedersen current in the ionosphere and via field aligned 
currents between the ionosphere and the magnetodisk. The strong field aligned currents accelerate electrons to very high speeds towards the ionosphere, generating the
main auroral emission (which thus maps to the position of the corotation breakdown).
This phenomenon was first explained by \citet{Bunce2001,Cowley2001,Hill2001}.

Recently, \citet{Palmaerts2014} used HST UV observations of the Jovian main auroral emission and found a transient localized brightness enhancement of the main emission close to noon local-time (see Figure~\ref{fig:HST}).
This small scale feature was observed on both hemispheres. 
Statistical analysis showed that this enhancement was localized between 10:00~LT and 15:00~LT in the southern hemisphere and between 09:00~LT and 15:30~LT in the northern hemisphere.
\citet{Palmaerts2014} found that the brightness of this localized peak can be 4.6 times larger than the average main emission.
This localized peak in the main auroral emission is also present in global MHD simulation by \citet{Chane2013,Chane2017}. 
It is observed in all simulations and occurs naturally. In the present paper, we will analyze in details 
simulation~2 of \citet{Chane2017} in order to understand what is 
causing this localized peak in the main emission in the simulations.

\begin{figure}
\begin{center}
\noindent\includegraphics[width=17pc]{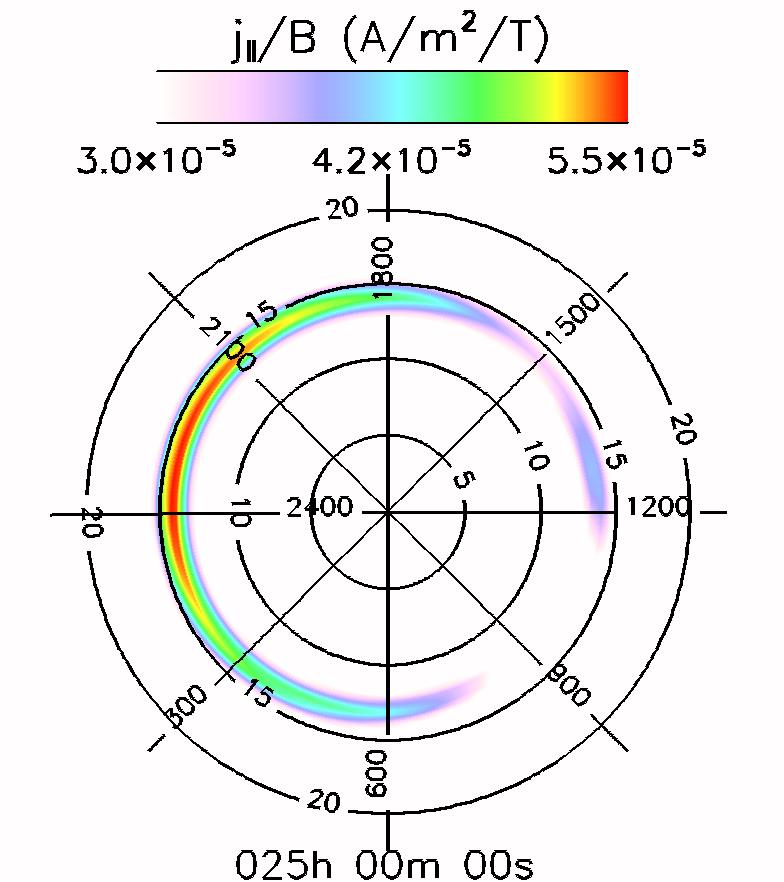}
\caption{$j_{\parallel}/B$ in the ionosphere (northern hemisphere) at $t=$25h00.
The localized peak is clearly visible around 13:00~LT.
The color bar is saturated, meaning that values below 3$\times$10$^{-5}$~A/m$^2$/T are shown in white, and
values above 5$\times$10$^{-5}$~A/m$^2$/T in red.
Local time and colatitude are overplotted in black. The Sun is located on the right.
}
\label{fig:aurorae2D}
\end{center}
\end{figure}

\section{Numerical Setup}
The numerical setup was already presented in details in \citet{Chane2017} (in the present paper, we simply re-analyze simulation~2). 
Nevertheless for the reader convenience, we give below a summary of this numerical setup.
Our simulations are global three dimensional one-fluid MHD simulations of the interaction between the solar wind and Jupiter's magnetosphere. 
We use the code MPI-AMRVAC to solve the equations \citep[see][]{Keppens2012,Porth2014,Xia2017}. The details of the model can be found in \citet{Chane2013}.
The equations solved in the model are the ideal MHD equations plus gravity. In addition, in an axisymmetric toroidal region representing the 
Io torus, a mass-loading source term is added to the equations. The total mass-loading in the torus is set to 1000~kg/s, and it is assumed that the 
neutral particles have a Keplerian velocity prior to ionization. 
The magnetosphere-ionosphere coupling is also introduced as a source term in the MHD equations, namely by including ion-neutral collisions in an axisymmetric 
region above the inner boundary. These collisions accelerate the plasma in the ionospheric region (thus initiating the rotation of the magnetosphere)
and allow for electrical current closure (since the collisions generate finite Pedersen and Hall conductivities in the ionospheric region).
It is assumed that the neutral particles in the ionosphere are rigidly corotating with Jupiter.

In our simulations, the inner boundary is located at 4.5~R$_{\textrm{\scriptsize{J}}}$ and the outer boundary at 189~R$_{\textrm{\scriptsize{J}}}$.
The inner boundary is not at 1~R$_{\textrm{\scriptsize{J}}}$, because the time-step is limited by the Courant-Friedrichs-Lewy (CFL) condition:
at 1~R$_{\textrm{\scriptsize{J}}}$ the Alfv\'en speed is so large that the time-steps would be too small to perform the simulations in a reasonable amount of time.
Note that in other global simulations of the Jovian magnetosphere published in the literature, the inner boundary is located even farther 
\citep[the closest being 8~R$_{\textrm{\scriptsize{J}}}$ in][]{Moriguchi2008}.
In our simulations, the ionospheric region is unrealistically large (between 4.5 and 8.5~R$_{\textrm{\scriptsize{J}}}$) because the numerical resolution is too coarse to simulate a smaller
ionospheric region and because increasing the numerical resolution would slow down the simulation too much.
Finally, in order to have a clear separation between the torus and the ionospheric region, the mass-loading cannot occur at the orbit of Io (5.9~R$_{\textrm{\scriptsize{J}}}$).
We therefore place the mass-loading in a torus of large radius 10~R$_{\textrm{\scriptsize{J}}}$ and of small radius 1~R$_{\textrm{\scriptsize{J}}}$ (thus between 9~R$_{\textrm{\scriptsize{J}}}$ 
and 11~R$_{\textrm{\scriptsize{J}}}$).
The drawback is that our model is not realistic close to the inner boundary (the ionosphere is too large and the Io torus is too far).
The main advantage is that the magnetosphere-ionosphere coupling occurs inside the numerical domain, above the inner boundary (not through the boundary like in other models),
precluding any spurious effect at the boundary to affect the coupling. This magnetosphere-ionosphere coupling has been extensively tested in \citet{Chane2013}, where it was shown to give 
realistic results, in agreement with both theories and observations.
Therefore, the aforementioned restrictions will not affect the results of the present study.

\begin{figure}
\begin{center}
\noindent\includegraphics[width=19pc]{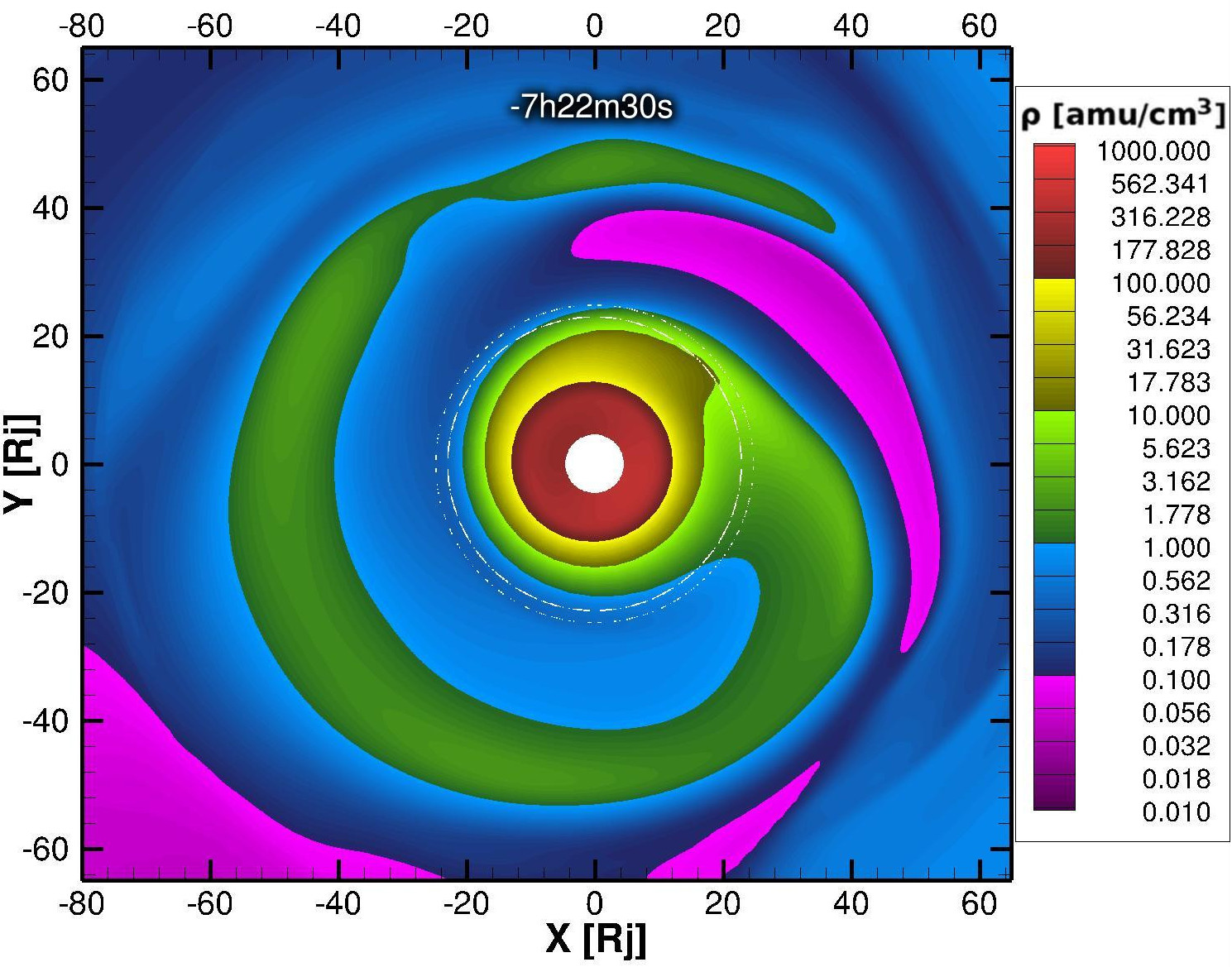}
\caption{Plasma density in the equatorial plane. Each order of magnitude is represented by a different color. 
The solar wind comes from the right. The high density regions display a spiral form.
}
\label{fig:spiral}
\end{center}
\end{figure}

\begin{figure*}
\begin{center}
\noindent\includegraphics[width=38pc]{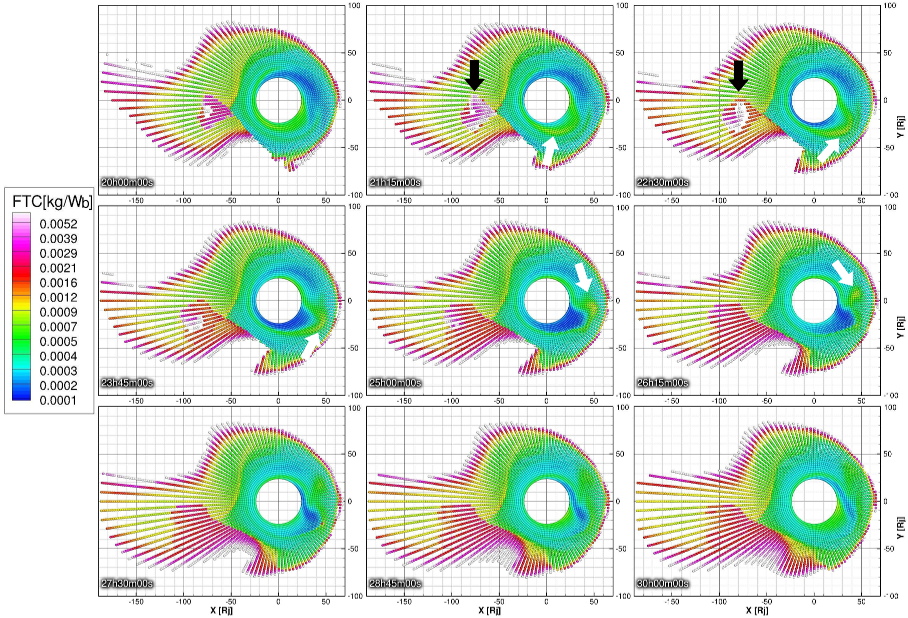}
\caption{Snapshots of flux tube content on closed field lines shown in the equatorial plane.
The solar wind comes from the right.
This figure highlights how heavy flux tubes can rapidly move radially away from Jupiter on the prenoon sector
when the centrifugal force increases (white arrows), and how plasmoids move away from Jupiter in the magnetotail at larger radial distances (black arrows). 
}
\label{fig:fluxTubes}
\end{center}
\end{figure*}

The solar wind parameters used in this simulation are the following: $v_x = -400$~km/s, $v_y=v_z=0$, $B_y=0.44$~nT, $B_x=B_z=0$, $T=15\,000$~K;
with $\rho = 0.162$~amu~cm$^{-3}$ for $t<$0 and then $\rho = 0.552$~amu~cm$^{-3}$ 
for $t\geqslant$0 (i.e. the time $t=$0 is defined as the instant when the high density solar wind reaches the magnetosphere).
Note that during the first 328 hours of the simulation (when $t<$0 and $\rho = 0.162$~amu~cm$^{-3}$ in the solar wind),
the simulation reaches a quasi steady state.
Although the outer boundary -- located at 189~R$_{\textrm{\scriptsize{J}}}$ -- is spherically shaped, the plasma density increase
(by a factor of 3.4) is introduced in a plane-parallel structure.
It should also be noted that the solar wind density is linearly increased during a time interval of one hour.
It means that at a given time, the imposed density at the outer boundary is not uniformed: it can be $0.162$~amu~cm$^{-3}$ for $x<x_1$,
$0.552$~amu~cm$^{-3}$ for $x>x_2$ (where $x_1$ and $x_2$ are the positions where the transition between low density and the high density solar wind occurs), 
and decrease linearly in between.
The simulations are performed on a static mesh, where three levels of refinement are used.
The smallest cells are 0.25~R$_{\textrm{\scriptsize{J}}}$ large and the largest cells 1~R$_{\textrm{\scriptsize{J}}}$ large.
The effective resolution is 800$\times$128$\times$128. This mesh is the same as the one used
for all the simulations in \citet{Chane2017}.
In the simulation, both the rotational and the magnetic axes are aligned with the z-axis.

\section{Localised peak and shearing motions}
\label{sec:setup}

In this section, we will show that the localized peak in our simulations is associated with shearing motions in the magnetodisk.
The central panels of Figure~\ref{fig:VrJpar} illustrate the periodic formation of this localized peak of electrical current 
in the ionosphere around noon local-time in all our simulations.
Every rotation period, the ionospheric field aligned currents are locally enhanced around noon.
In our simulations, the ionospheric field aligned current are strongest on the night side, and display
a minimum on the prenoon sector. This minimum corresponds to the main oval discontinuity observed by \citet{Radioti2008}.
As discussed by \citet{Chane2017}, the field aligned currents are stronger on the night-side because
the field lines are more elongated at that location. As a result, this is also where azimuthal bending of the field lines due to subcorotating plasma
generates the largest radial electrical currents (which are closed in the ionosphere via field aligned currents).
Figure~\ref{fig:VrJpar} shows that, in our simulation, $j_{\parallel}/B$ close to noon increases by 60\% within 5~hours
before returning to lower values 5~hours later.
Figure~\ref{fig:aurorae2D} displays $j_{\parallel}/B$ in our simulation, in the ionosphere when the localized peak 
is present (at time $t=$25h00). The main emission discontinuity (between 8:00~LT and 15:00~LT) and the localized peak (around 13:00~LT)
are clearly visible in this figure.

The left panels of Figure~\ref{fig:VrJpar} show that in our simulations, the periodic localized peak in ionospheric
field aligned current is associated with an inward motion of the plasma on the day-side magnetodisk.
At time $t=$20h00, there are almost no inward motions of plasma visible inside the magnetodisk, and the localized 
ionospheric electrical current peak is absent. 
At time $t=$22h30, a region of inward moving plasma becomes visible between 25 and 50~R$_{\textrm{\scriptsize{J}}}$
on the prenoon sector. One can also see on the central panel that this corresponds to the beginning of the formation of 
the localized peak in the ionosphere.
At time $t=$25h00 the ionospheric peak is clearly visible, while the area of inward plasma motion is now larger in the magnetodisk,
and the plasma moves faster inwards.
From this moment on, the portion of the magnetodisk displaying strong negative radial velocities decreases in size, while the ionospheric peak
becomes less visible and disappears almost completely at time $t=$30h00.
This phenomenon occurs every rotation period.
Note that solar wind ram pressure, and thus the size of the magnetosphere affects the localized peak. 
Namely, the peak is more visible when the solar wind ram pressure is high (not shown here). This is because: 1) more currents are present in the peak, and 
2) the background main emission currents are weaker 
\citep[since the main emission discontinuity is more pronounced when the solar wind ram pressure is high, see][]{Chane2017},
leading to a greater contrast with the localized peak. 
It should also be noted that the localized peak is not generated by the solar wind density increase at time t$=0$.
The peak is also clearly visible in simulation~1 of 
\citet[][]{Chane2017} (not shown here) although the solar wind density imposed at the outer boundary did not change during this simulation.

The ratio between the plasma azimuthal velocity and the theoretical rigid corotation velocity is displayed in the
right panels of Figure~\ref{fig:VrJpar}.
One can clearly see that the plasma rotates faster at dawn than at dusk in our simulation \citep[note that this is also seen in measurements from the Galileo spacecraft, see][]{Woch2004,Krupp2004}.
These panels also show that between $t=$20h00 and $t=$30h00, on the pre-noon sector, at large radial distances, close to the magnetopause, 
a channel of fast rotating plasma develops in the simulation. This channel however, does not seem to be causing the localized ionospheric current peak.
This is particularly visible at time $t=$30h00, where the channel is clearly present but the localized current peak is not.
This channel is caused by depleted flux tubes at large radial distances that were blocked between the heavy magnetodisk and the magnetopause.
Once the heavy flux tubes move inward close to noon (see left panels) the rotation of the depleted flux tubes is not hindered any more and the plasma is accelerated 
\citep[for more details, see][]{Southwood2016,Radioti2017}.

The shearing motions produce field aligned currents that close in the ionosphere via two different mechanisms.
First, the inward motion of the plasma tends to reduce the subcorotarion of the plasma and hence decreases the radial current, 
conversely plasma moving outward produces more radial current. 
If $r^2 j_r$ decreases with radial distance (since $\nabla \cdot \mathbf{j} = 0$), part of the radial electrical current is converted in azimuthal currents or in field aligned currents. 
This mechanism mostly generates field aligned currents on the inner region of the 
negative radial velocity area (the dark blue region inside the magnetodisk in the left panels of Figure~\ref{fig:VrJpar}).
The second mechanism involves azimuthal currents closing in the ionosphere: on the one hand plasma moving outward stretches the closed magnetic field lines of Jupiter, which
increases the azimuthal electrical current. On the other hand, inward plasma motions compress the field lines and less current is generated.
The azimuthal gradient of azimuthal electrical current produces field aligned currents that are closed in the ionosphere. 
This second mechanism mainly creates field aligned currents on the eastern part of the negative radial velocity area.
As a result of these two mechanisms, field aligned currents are mostly produced on the inner eastern part of the negative radial velocity area (dark blue area in the left panels of Figure~\ref{fig:VrJpar}).

\begin{figure*}
\begin{center}
\noindent\includegraphics[width=38pc]{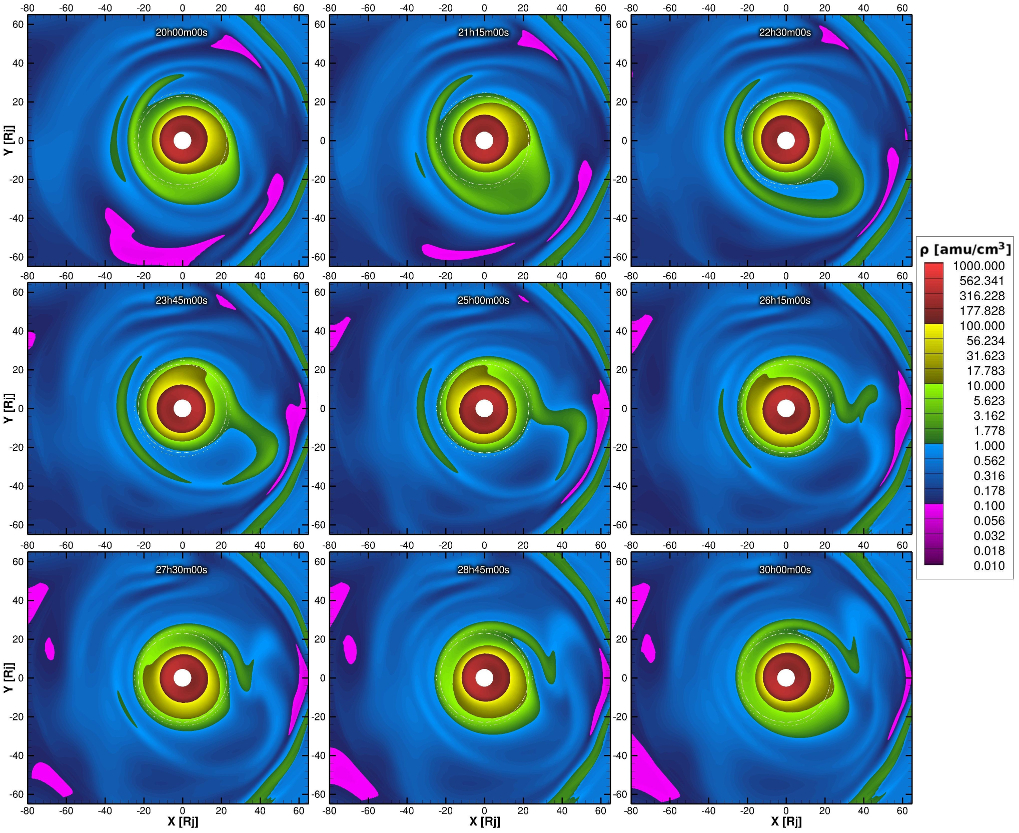}
\caption{Snapshots of the plasma density in the equatorial plane. The solar wind comes from the right.
It shows how a high density area moves radially away in the pre-noon sector prior to 25h00 and how
it then falls back at lower radial distances in the post-noon sector.
}
\label{fig:densityMovie}
\end{center}
\end{figure*}

\section{Radial motion of heavy flux tubes}
\label{sec:magneto}

\begin{figure*}
\begin{center}
\noindent\includegraphics[width=38pc]{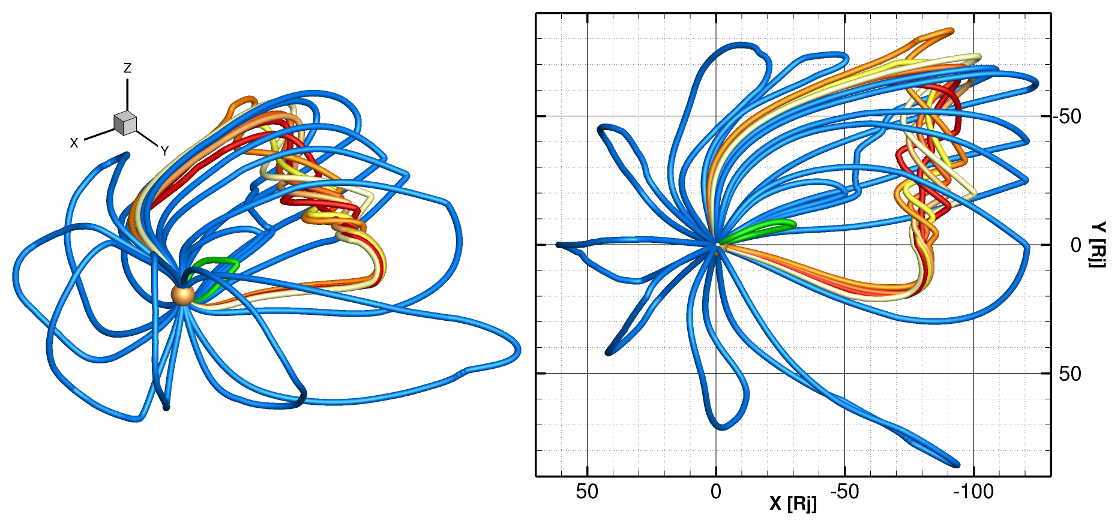}
\caption{Selected magnetic field lines at $t=$23h45. A plasmoid ejection is visible on the night-side (reddish and yellowish colors).
A field line emanating from the high density spiral is plotted in green. The other field lines are shown in blue. The solar wind comes from the left.
}
\label{fig:plasmoid}
\end{center}
\end{figure*}

In the present section, we will seek to understand why the presence of the localized ionospheric current peak occurs once 
per rotation period in our simulations.
First, one needs to understand that in our simulations, the plasma density in the magnetodisk is far from axisymmetric, even though the Iogenic 
mass-loading is axisymmetric. This feature is illustrated in Figure~\ref{fig:spiral} where it can be seen that the dense plasma region assume a 
spiral shape \citep[as explained in][]{Chane2017}. This spiral rotates with Jupiter. 
As we will see below, the spiral is not always as clearly defined and as long as in Figure~\ref{fig:spiral}, but it is always present.
As a result, in our simulations at any given point inside the magnetodisk, the plasma density increases significantly each rotation period.
It should be noted that it is not clear whether the unique spiral arm shown in Figure~\ref{fig:spiral} would be affected by the resolution
of the simulation. It might be that a simulation with an higher numerical resolution would produce more than one spiral arm.

As seen in Figure~\ref{fig:VrJpar}, the plasma azimuthal velocity is higher at dawn. 
These high azimuthal velocities produce a stronger centrifugal force and the radial velocity is therefore, as can be 
seen in the figure, also higher at dawn.
Every rotation period, heavy loaded flux tubes arrives at dawn and are accelerated. The centrifugal force then exceeds the sum of the inward pointing forces (mostly
the magnetic tension), and the heavy flux tubes rapidly move away from Jupiter. 
This process is clearly visible in Figure~\ref{fig:fluxTubes}, where the flux tube content is plotted in the equatorial plane.
At time $t=$21h15, the heavy flux tubes at about 30~R$_{\textrm{\scriptsize{J}}}$ reach dawn and start to rapidly move outward.
At time $t=$22h30, the heavy flux tubes are now around 10:00LT and at a radial distance of $r\simeq$~45~R$_{\textrm{\scriptsize{J}}}$.
When the heavy flux tubes reach $\sim$50~R$_{\textrm{\scriptsize{J}}}$ at time $t=$25h00, they are almost at noon local-time.
In this area the azimuthal velocities are lower 
\citep[mostly because field lines are more dipolar and the magnetic tension in the azimuthal direction is thus weaker, see][]{Chane2017}.
As a result, the centrifugal force is weaker and the heavy flux tubes begin to slowly move inward (this negative radial velocity is clearly visible in Figure~\ref{fig:VrJpar}).
At time $t=$26h15, the heavy flux tubes just passed noon local-time ($\sim$12:30LT) and are now located at a radial distance of $r\simeq$~43~R$_{\textrm{\scriptsize{J}}}$.
This periodic process occurs once per rotation period and produces the shearing motion responsible for the localized peak of ionospheric electrical currents in our simulations.
Note that if the high density spiral shown in Figure~\ref{fig:spiral} had more than one arm (which might have been the case if an higher 
numerical resolution had been used), the localized peak would be present more than once per rotation period.

The equatorial plasma density is shown in Figure~\ref{fig:densityMovie}. Between $t=$21h15 and 22h30, one can see how the high density spiral enters the region
of large azimuthal velocity, and how it deforms due to the radial motions caused by the higher centrifugal force.
At time $t=$23h45 a high density island (in green) seems to detach from the spiral on the night side. This is not a plasmoid ejection (see Figure~\ref{fig:plasmoid}) but simply an artifact of our color bar.
Since we use a color bar where each order of magnitude is displayed with a different color, this green island has no physical meaning: it simply means that the plasma density is higher
than 1~amu/cm$^3$ in this green island and that from $t=$23h45 a part of the spiral is now displayed in light blue.
At time $t=$25h00, at noon local-time, the outer part of the spiral is rotating faster than the inner part (see Figure~\ref{fig:VrJpar}) which results in a deformation of the spiral shape.
Between $t=$25h00 and 27h30, one can clearly see in Figure~\ref{fig:densityMovie} that the high density region around noon local-time is moving inward.


In our simulations, plasmoids are episodically (but not periodically) ejected, releasing large quantities of plasma in the magnetotail.
Figure~\ref{fig:plasmoid} shows such a plasmoid at time $t=$23h45. The plasmoid is located between $\sim$70 and $\sim$100~R$_{\textrm{\scriptsize{J}}}$ downtail and
has an azimuthal extend of about 45$^\circ$ (between midnight and 3:00~LT).
Figure~\ref{fig:plasmoid} also shows a field line (in green) crossing the high density spiral shown in Figure~\ref{fig:densityMovie}.
This field line is crossing the equatorial plane in the middle of the green island shown in Figure~\ref{fig:densityMovie}.
The field line is mostly dipolar and is clearly not part of a plasmoid ejection. 
The plasmoid ejections occur at larger radial distances and
can be easily spotted when looking at the flux-tube content (see Figure~\ref{fig:fluxTubes}). This is because: 1) field lines from a plasmoid are much longer than 
dipolar closed field lines, and 2) because a large amount of plasma is usually ejected within the plasmoid.
This is, for instance, particularly true on the first panels of Figure~\ref{fig:fluxTubes} where a plasmoid can easily be spotted in the magnetotail.

It is important to understand that in our simulations, even though the localized current peak in the ionosphere is associated with plasma rapidly moving
away from the planet, it is not related to plasmoid ejections. This is because these ejections occur at larger radial distance and on the night side.

\section{Comparison with the observations}
\label{sec:CompWithObs}

\begin{figure}
\begin{center}
\noindent\includegraphics[width=19pc]{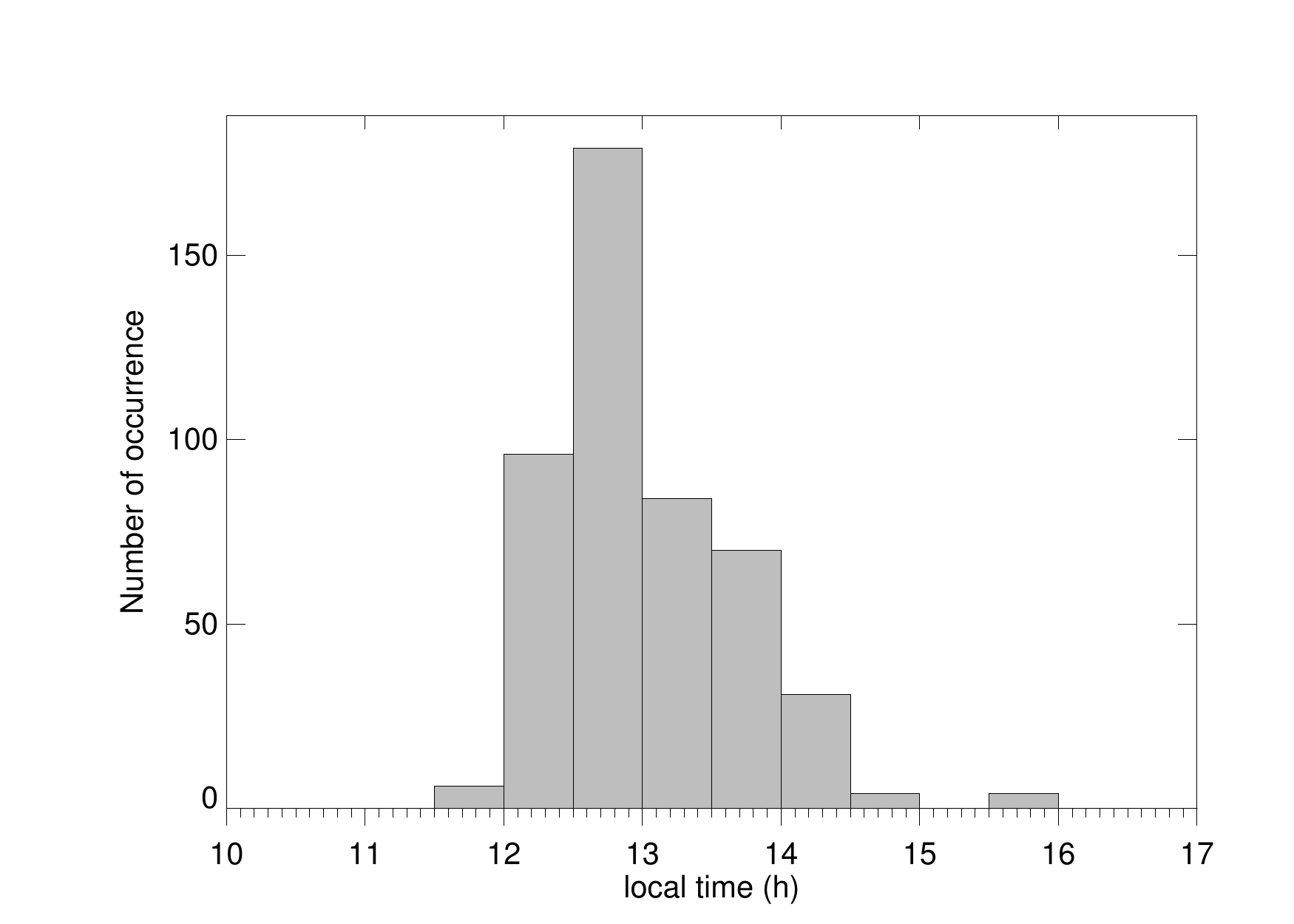}
\caption{
Distribution of the local time position of the localized peak during 133 hours of simulations. }
\label{fig:distribution}
\end{center}
\end{figure}

As explained in the previous section, the localized peak in the global MHD simulations results from an inward motion of plasma in the noon local time sector.
Although the dayside magnetosphere has not been widely sampled by the Galileo spacecraft, such inward flows have been detected in the equatorial plane 
around noon by the plasma instrumentation on board \citep{Frank2004,Waldrop2015}. 
The peak in the simulated field aligned currents leads to a localized brightening of the main auroral emission which is thought 
to be the transient auroral spot in the main emission depicted by \citet{Palmaerts2014}. As shown in the histogram in Figure~\ref{fig:distribution}, 
the localized peak in the simulations is mostly found between 12:00 and 14:00~LT, similarly to the local time distribution of the observed auroral spot \citep{Palmaerts2014}.
In both simulations and auroral observations, the shape of the peak evolves with time. The localized peak described in the present study is somewhat 
broader than the auroral spot (2-3~h in LT compared to around 1~h in LT), but this might be due to the limited numerical resolution of the simulations. 
Finally, the localized peak is present 54\% of the time during the 133~hours of simulations considered in this study, which is close to the occurrence 
probability of the auroral spot of 60\% inferred by \citet{Palmaerts2014} using 1685 HST auroral images. It has been shown in the previous section that 
the peak appears once per planetary rotation in our simulations. Such an occurrence period cannot be inferred from the HST observations since the observing 
sequences last only $\sim$45 min and several days can separate two consecutive sequences. 
The similarities between the localized peak in the simulations and the transient auroral spot in the main emission lead to the conclusion that the auroral 
spot might be generated by the process described in this manuscript, namely an inward motion of flux tubes in the local noon sector. 
Note that \citet{Palmaerts2014} already suggested that the shear caused by intermittent inward plasma motions in the magnetodisk near noon
may be responsible for the transient small-scale structure that are observed in the main auroral emission.

Recently, \citet{Nichols2017} presented HST observations of an auroral spot within the main emission located at 13:00-14:00~LT very similar to the one observed
by \citet{Palmaerts2014}. They also showed that the intensity of this localized spot oscillated with a period of approximately 10 minutes. 
They suggested that these oscillations were generated by waves confined in the low density region of the magnetosphere situated between the ionosphere and the plasma sheet.
Unfortunately, we do not know whether, in the simulations, the localized auroral peak also oscillates with a period of about 10 minutes. This is because the output cadence
was set to 7.5 minutes in these simulations: saving the simulation outputs every 30 seconds would have generated an extremely large amount of data and we did not see the 
need to do so when we performed the simulations, before \citet{Nichols2017} observations. We will make sure to use a higher cadence in future runs, to verify whether the
oscillations are also present in our simulations and to understand what is causing them.

\section{Summary and Conclusions}
\label{sec:conclusions}

The localized peak in the intensity of the main auroral emission observed by \citet{Palmaerts2014} is also present in all our global MHD
simulations. In the simulations, this peak occurs periodically---once every rotation period---and is caused by shearing motions in the 
magnetodisk on the day-side.
A high density spiral, which rotates around Jupiter and is mostly confined close to the equatorial plane, is always present in our simulations.
Note that the number of spiral arms, and thus the periodicity of the localized peak might be affected by the numerical resolution.
When the high density plasma reaches---every rotation period---high azimuthal speed at dawn (in the Jovian magnetosphere, the plasma rotates faster at dawn than at dusk)
the centrifugal force is so strong than the flux-tubes rapidly move to larger radial distances. When these flux-tubes reach noon, where the azimuthal velocity is lower
and the centrifugal force is thus weaker, they 
fall back closer to Jupiter, generating the aforementioned shearing motions responsible for the localized peak in the main emission.


{\scriptsize
{\bf Acknowledgments.}
E.C. was funded by the Research Foundation-Flanders (grant FWO 12M0115N).
B.P. was supported by the PRODEX program managed by ESA in collaboration with the Belgian Federal Science Policy Office.
A.R. was funded by the Belgian Fund for Scientific Research (FNRS).
Computations were performed
on the supercomputers Muk and ThinKing (provided by the VSC (Flemish Supercomputer Center), funded by 
the Research Foundation - Flanders (FWO) and the Flemish Government - department EWI), 
as well as on Trillian (a Cray XE6m-200 supercomputer at the University of New Hampshire supported by the NSF MRI program under grant PHY-1229408).
E.C. would also like to acknowledge the ISSI teams ``Coordinated Numerical Modeling of the Global Jovian and Saturnian Systems'' 
and ``How does the Solar Wind Influence the Giant Planet Magnetospheres?'' for useful discussions.
}



\bibliographystyle{elsarticle-harv}
\bibliography{biblio}

\begin{thebibliography}{21}
\expandafter\ifx\csname natexlab\endcsname\relax\def\natexlab#1{#1}\fi
\expandafter\ifx\csname url\endcsname\relax
  \def\url#1{\texttt{#1}}\fi
\expandafter\ifx\csname urlprefix\endcsname\relax\def\urlprefix{URL }\fi

\bibitem[{{Badman} et~al.(2015){Badman}, {Branduardi-Raymont}, {Galand},
  {Hess}, {Krupp}, {Lamy}, {Melin}, and {Tao}}]{Badman2015}
{Badman}, S.~V., {Branduardi-Raymont}, G., {Galand}, M., {Hess}, S.~L.~G.,
  {Krupp}, N., {Lamy}, L., {Melin}, H., {Tao}, C., Apr. 2015. {Auroral
  Processes at the Giant Planets: Energy Deposition, Emission Mechanisms,
  Morphology and Spectra}. Space Sci. Rev. 187, 99--179.

\bibitem[{{Bunce} and {Cowley}(2001)}]{Bunce2001}
{Bunce}, E.~J., {Cowley}, S.~W.~H., Aug. 2001. {Divergence of the equatorial
  current in the dawn sector of Jupiter's magnetosphere: analysis of Pioneer
  and Voyager magnetic field data}. Planetary and Space Science 49, 1089--1113.

\bibitem[{{Chan{\'e}} et~al.(2017){Chan{\'e}}, {Saur}, {Keppens}, and
  {Poedts}}]{Chane2017}
{Chan{\'e}}, E., {Saur}, J., {Keppens}, R., {Poedts}, S., Feb. 2017. {How is
  the Jovian main auroral emission affected by the solar wind?} Journal of
  Geophysical Research (Space Physics) 122, 1960--1978.

\bibitem[{{Chan{\'e}} et~al.(2013){Chan{\'e}}, {Saur}, and
  {Poedts}}]{Chane2013}
{Chan{\'e}}, E., {Saur}, J., {Poedts}, S., May 2013. {Modeling Jupiter's
  magnetosphere: Influence of the internal sources}. Journal of Geophysical
  Research (Space Physics) 118, 2157--2172.

\bibitem[{{Clarke} et~al.(2004){Clarke}, {Grodent}, {Cowley}, {Bunce}, {Zarka},
  {Connerney}, and {Satoh}}]{Clarke2004}
{Clarke}, J.~T., {Grodent}, D., {Cowley}, S.~W.~H., {Bunce}, E.~J., {Zarka},
  P., {Connerney}, J.~E.~P., {Satoh}, T., 2004. {Jupiter's aurora}. In:
  {Bagenal, F., Dowling, T.~E., \& McKinnon, W.~B.} (Ed.), Jupiter.~The Planet,
  Satellites and Magnetosphere. pp. 639--670.

\bibitem[{{Cowley} and {Bunce}(2001)}]{Cowley2001}
{Cowley}, S.~W.~H., {Bunce}, E.~J., Aug. 2001. {Origin of the main auroral oval
  in Jupiter's coupled magnetosphere-ionosphere system}. Planetary and Space
  Science 49, 1067--1088.

\bibitem[{Frank and Paterson(2004)}]{Frank2004}
Frank, L., Paterson, W., 2004. Plasmas observed near local noon in jupiter's
  magnetosphere with the galileo spacecraft. Journal of Geophysical Research:
  Space Physics 109~(A11).

\bibitem[{{Grodent}(2015)}]{Grodent2015}
{Grodent}, D., Apr. 2015. {A Brief Review of Ultraviolet Auroral Emissions on
  Giant Planets}. Space Sci. Rev. 187, 23--50.

\bibitem[{{Hill}(2001)}]{Hill2001}
{Hill}, T.~W., May 2001. {The Jovian auroral oval}. J. Geophys. Res. 106,
  8101--8108.

\bibitem[{{Keppens} et~al.(2012){Keppens}, {Meliani}, {van Marle}, {Delmont},
  {Vlasis}, and {van der Holst}}]{Keppens2012}
{Keppens}, R., {Meliani}, Z., {van Marle}, A.~J., {Delmont}, P., {Vlasis}, A.,
  {van der Holst}, B., Feb. 2012. {Parallel, grid-adaptive approaches for
  relativistic hydro and magnetohydrodynamics}. Journal of Computational
  Physics 231, 718--744.

\bibitem[{{Krupp} et~al.(2004){Krupp}, {Vasyliunas}, {Woch}, {Lagg}, {Khurana},
  {Kivelson}, {Mauk}, {Roelof}, {Williams}, {Krimigis}, {Kurth}, {Frank}, and
  {Paterson}}]{Krupp2004}
{Krupp}, N., {Vasyliunas}, V.~M., {Woch}, J., {Lagg}, A., {Khurana}, K.~K.,
  {Kivelson}, M.~G., {Mauk}, B.~H., {Roelof}, E.~C., {Williams}, D.~J.,
  {Krimigis}, S.~M., {Kurth}, W.~S., {Frank}, L.~A., {Paterson}, W.~R., 2004.
  {Dynamics of the Jovian magnetosphere}. In: {Bagenal, F., Dowling, T.~E., \&
  McKinnon, W.~B.} (Ed.), Jupiter.~The Planet, Satellites and Magnetosphere.
  pp. 617--638.

\bibitem[{{Moriguchi} et~al.(2008){Moriguchi}, {Nakamizo}, {Tanaka}, {Obara},
  and {Shimazu}}]{Moriguchi2008}
{Moriguchi}, T., {Nakamizo}, A., {Tanaka}, T., {Obara}, T., {Shimazu}, H., May
  2008. {Current systems in the Jovian magnetosphere}. J. Geophys. Res. (Space
  Physics) 113, A05204.

\bibitem[{Nichols et~al.(2017)Nichols, Yeoman, Bunce, Chowdhury, Cowley, and
  Robinson}]{Nichols2017}
Nichols, J.~D., Yeoman, T.~K., Bunce, E.~J., Chowdhury, M.~N., Cowley, S.
  W.~H., Robinson, T.~R., 2017. Periodic emission within jupiter's main auroral
  oval. Geophysical Research Letters.
\newline\urlprefix\url{http://dx.doi.org/10.1002/2017GL074824}

\bibitem[{Palmaerts et~al.(2014)Palmaerts, Radioti, Grodent, Chan{\'e}, and
  Bonfond}]{Palmaerts2014}
Palmaerts, B., Radioti, A., Grodent, D., Chan{\'e}, E., Bonfond, B., 2014.
  Transient small-scale structure in the main auroral emission at jupiter.
  Journal of Geophysical Research: Space Physics 119~(12), 9931--9938.

\bibitem[{{Porth} et~al.(2014){Porth}, {Xia}, {Hendrix}, {Moschou}, and
  {Keppens}}]{Porth2014}
{Porth}, O., {Xia}, C., {Hendrix}, T., {Moschou}, S.~P., {Keppens}, R., Sep.
  2014. {MPI-AMRVAC for Solar and Astrophysics}. ApJS 214, 4.

\bibitem[{{Radioti} et~al.(2008){Radioti}, {G{\'e}rard}, {Grodent}, {Bonfond},
  {Krupp}, and {Woch}}]{Radioti2008}
{Radioti}, A., {G{\'e}rard}, J., {Grodent}, D., {Bonfond}, B., {Krupp}, N.,
  {Woch}, J., Jan. 2008. {Discontinuity in Jupiter's main auroral oval}.
  Journal of Geophysical Research (Space Physics) 113, A01215.

\bibitem[{Radioti et~al.(2017)Radioti, Grodent, G{\'e}rard, Southwood,
  Chan{\'e}, Bonfond, and Pryor}]{Radioti2017}
Radioti, A., Grodent, D., G{\'e}rard, J.-C., Southwood, D.~J., Chan{\'e}, E.,
  Bonfond, B., Pryor, W., 2017. Stagnation of saturn's auroral emission at
  noon. Journal of Geophysical Research: Space Physics 122~(6), 6078--6087,
  2016JA023820.

\bibitem[{Southwood and Chan{\'e}(2016)}]{Southwood2016}
Southwood, D.~J., Chan{\'e}, E., 2016. High-latitude circulation in giant
  planet magnetospheres. Journal of Geophysical Research: Space Physics
  121~(6), 5394--5403, 2015JA022310.

\bibitem[{Waldrop et~al.(2015)Waldrop, Roelof, and Fritz}]{Waldrop2015}
Waldrop, L., Roelof, E., Fritz, T., 2015. Three-dimensional convective flows of
  energetic ions in jupiter's equatorial magnetosphere. Journal of Geophysical
  Research: Space Physics 120~(12).

\bibitem[{{Woch} et~al.(2004){Woch}, {Krupp}, {Lagg}, and
  {Tom{\'a}s}}]{Woch2004}
{Woch}, J., {Krupp}, N., {Lagg}, A., {Tom{\'a}s}, A., Jan. 2004. {The structure
  and dynamics of the Jovian energetic particle distribution}. Advances in
  Space Research 33, 2030--2038.

\bibitem[{Xia et~al.(2018)Xia, Teunissen, El~Mellah, Chan{\'e}, and
  Keppens}]{Xia2017}
Xia, C., Teunissen, J., El~Mellah, I., Chan{\'e}, E., Keppens, R., 2018.
  Mpi-amrvac 2.0 for solar and astrophysical applications. The Astrophysical
  Journal Supplement Series 234~(2), 30.

\end{thebibliography}







\end{document}